\documentclass[12pt]{article} 
\usepackage{graphicx}  
\begin{document}
\title{\bf{How to test the gauge invariant \\ non local quantum dynamics of the Aharonov-Bohm effect}}
\author{T. Kaufherr$^{1,2}$}
\maketitle
\noindent$^{1}$\emph{School of Physics and Astronomy,Tel Aviv University,Tel Aviv 69978, Israel}\\
$^{2}$\emph{E-mail: trka@post.tau.ac.il}
\maketitle
\begin{abstract}
The gauge invariant non local quantum dynamics of the Aharonov-Bohm effect can be tested experimentally by measuring the
instantaneous shift of the velocity distribution occurring when the particle passes by the flux line. It is shown that in relativistic quantum theory it is possible to measure the instantaneous velocity with accuracy sufficient to detect the change of the velocity distribution. In non relativistic quantum theory the instantaneous velocity can be measured to any desired accuracy.
\end{abstract}
\section{Introduction}
\label{intro}
In a previous article\cite{Aharonov:Kaufherr} it has been shown, that the shift of the interference pattern in the Aharonov-Bohm [AB] effect\cite{Aharonov:Bohm}, is due to a nonlocal dynamical exchange of a gauge invariant quantity between the charged particle and the flux line. On a basic level, this relates to the question concerning the role of the electromagnetic potentials in quantum theory. Obviously, the non-gauge-invariant potentials are an auxiliary mathematical tool, with no physical significance.
The quantity in  question is modulo velocity or $\langle\cos\frac{mv_{y}L}{\hbar} \rangle$, where $L$ is the separation between the particle's wave packets, see figure 1.
\begin{figure}
\centering
\includegraphics{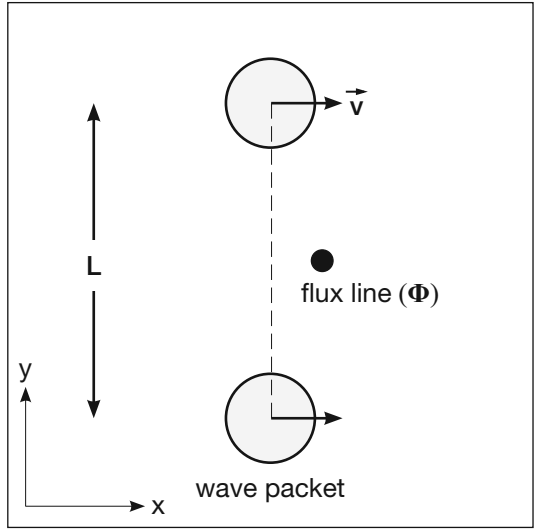}
\caption{The exchange occurs instantly, when the (imaginary) line connecting the two wave packets crosses the flux line.}
\end{figure}

The exchange occurs instantly, when the (imaginary) line connecting the two wave packets crosses the flux line (if one ignores the time it takes the wave packets to pass by the flux line). It also occurs non locally, since the wave packets are separated and have not completed a closed loop.
There arises the question, how to translate this into an observable effect?  The problem is, that one cannot measure directly functions of momentum such as $\cos\frac{pL}{\hbar}$ since it
violates the continuity equation.  Thus, one considers the particle's velocity distribution $P(v_{y})$. One can show that its Fourier transform, which is directly related to $\langle\cos\frac{mv_{y}L}{\hbar} \rangle$, changes. By comparison, none of the moments of momentum $\langle p_{y}^{n} \rangle$ changes. 
However, in order to demonstrate the instantaneous change of the velocity distribution, it is necessary that the velocity may be measured in as short a time as desired. Below, it is shown that the time needed to measure velocity tends to zero in non relativistic theory. In relativistic theory there is a limitation, and velocity cannot be measured in a time shorter than $\frac{L}{c}$. Still, to a good approximation, this may be considered an instantaneous measurement. Since a measurement of velocity does not change the velocity, an observed shift of the velocity distribution will attest that the change existed prior to the measurement, i.e., shortly after the wave packets passed by the flux line.
\paragraph{}It is also shown that the application of von Neumann's  measurement theory\cite{von:Neumann} to the measurement of velocity\cite{Aharonov:Safko} is straightforward, although the theory applies to canonical variables, while velocity is a non canonical variable.
 In particular, a measurement of velocity which involves two successive position measurements is considered. It seems, that in quantum theory such a measurement is impossible, for already the first position  measurement changes the  velocity in an uncertain way. However, by applying an additional interaction to the measuring device\cite{Bohr:Rosenfeld}\cite{Aharonov:Safko}, it is possible to cancel the effect of the measurement on its \emph{recording}, so that it monitors the distance that the \emph{unmeasured particle} would have travelled during the time of  measurement.
\paragraph{} After a brief introduction of von Neumann's theory, its application to the measurement of position$(\ref{sec:2.1})$ and to the measurement of velocity$(\ref{sec:2.2})$, the measurement of velocity by means of two position measurements is described$(\ref{sec:2.3})$. In section $(\ref{sec:3})$, the AB case is addressed. In section $(\ref{sec:4})$, two Gedanken experiments are described.
\section{Measurement of velocity}
\label{sec:2}
\subsection{von Neumann's measurement theory, measurement of position}
\label{sec:2.1}
Von Neumann's central claim was that, not only in classical physics but also in non relativistic quantum theory, any canonical
dynamical variable may be measured to any required accuracy and in as short a time as desired,  \emph{without changing it}, and that this is
achieved by using the coupling 
\begin{equation}
H_{I}=g(t)A_{S}B_{MD}\,,
\label{eq:II:I:1}
\end{equation}
where $A_{S}$ and $B_{MD}$ are canonical dynamical variables that pertain to the observed system and measuring device, respectively. The quantity $A_{S}$
 is the object of the  measurement. $g(t)$ is a gate function:
$g(t)=g_{0}$ in the interval $[0,T_{0}]$
and equals to zero otherwise.
It is assumed that $T_{0}\rightarrow 0$, while
$G=\int_{0}^{T_{0}} g(t)dt=g_{0}T_{0}$
remains finite and, specifically, that $G\geq 1$. Thus, $g(t)$ effects an \emph{impulsive measurement}. Note that $(\!\!~\ref{eq:II:I:1})$
is a generalization of the Stern Gerlach interaction.
\paragraph{}Rather than continuing in general terms, let us consider
 the measurement of a particle's instantaneous position $y(t=0)$. In this case, the interaction is given by $H_{I}=g(t)yq$
where $q$ is the coordinate of the measuring device. The overall Hamiltonian is then $H=\frac{p_{y}^{2}}{2m}+\frac{\pi^{2}}{2M}+g(t)yq$. $m$ and $M$ are the masses of the observed particle and the measuring device, and the momenta $p_{y}$ and $\pi$ are canonically conjugate to the coordinates $y$ and $q$, respectively. It is assumed that the measuring device is very heavy so that its kinetic energy term may be ignored. We thus consider the  Hamiltonian
\begin{equation}
H=\frac{p_{y}^{2}}{2m}+g(t)yq\,.
\label{eq:II:I:2}
\end{equation}
The most basic requirement from a measurement is, that there exists a variable of the measuring device, the "pointer", whose change during the measurement is proportional to the quantity that we want to measure. In the present case, $\pi$ is the pointer. We have $\pi(t\geq T_{0})=\pi(0)+\int_{0}^{T_{0}}g(t)y(t)dt$\cite{footnote}.
After the measurement, the pointer's displacement $\delta\pi(t)=\pi(t)-\pi(0)$ is given by
\begin{equation}
\delta\pi(t\geq T_{0})=G\left[y(0)+\frac{p_{y}(0)}{m}\frac{T_{0}}{2}+\frac{Gq(0)}{m}\frac{T_{0}}{6}\right]\,.
\label{eq:II:I:3}
\end{equation}
A result of a measurement is meaningful only if the pointer's displacement is much larger than the error, or, if
\begin{equation}
\delta\pi(t\geq T_{0}) \gg \Delta\pi(0)\,.
\label{eq:II:I:4}
\end{equation}
It is also assumed that the pointer is sharp, i.e., the uncertainty $\Delta\pi(0)$ is small. Thus, for a good measurement
of the position $y(0)$ it is necessary that
\begin{equation}\label{eq:II:I:5}
Gy(0) \gg \Delta\pi(0)\,,
\end{equation}
as well as that the additional third term in the square brackets of $(\ref{eq:II:I:3})$ may be ignored, i.e.,
\begin{equation}\label{eq:II:I:6}
\frac{G^{2}\Delta q(0)}{m}\frac{T_{0}}{6}\leq \Delta\pi(0)\,.
\end{equation}
During the measurement, the observed particle is accelerated by a large, uncertain amount, $\Delta \ddot{y}=\frac{ g(t)\Delta q(0)}{m}\geq\frac{ \hbar g(t)}{m\Delta\pi(0)}$, which is $\sim \frac{g(t)}{\Delta\pi(0)}$ for an electron. Thus, $(\ref{eq:II:I:6})$ means that this
 acceleration may be ignored at the impulsive limit.
Alternatively, one can assume that the kinetic energy is negligible compared to the interaction.
Either way, the particle's position does not change perceptibly during the time of measurement, which confirms von Neumann's claim.
\paragraph{}To conclude, the conditions $(\ref{eq:II:I:5})$ and $(\ref{eq:II:I:6})$ guarantee that the particle's instantaneous position at $t=0$ is being measured. However, after the measurement, the particle's velocity
$v_{y}(t\geq T_{0})=\frac{p_{y}(0)}{m}+\frac{Gq(0)}{m}$ is very uncertain.
\paragraph{}Note that to avoid the relativistic domain, the uncertain velocity must satisfy $\frac{G\Delta q(0)}{m}\ll c$, from which it follows that the uncertainty of the measured position $\Delta y\gg \frac{\hbar}{mc}=\lambda_{c}$, where $\lambda_{c}$ is the particle's Compton wavelength. The duration of the measurement is limited as well, by $T_{0}\gg\frac{h}{ mc^{2}}$, $\sim 10^{-20}$ sec for an electron. In non relativistic theory defined by $c\rightarrow \infty$, the instantaneous position may be measured to any desired accuracy and in as short a time as one may wish.
\subsection{The applicability of von Neumann's theory to the Measurement of velocity}
\label{sec:2.2}
Von Neumann's measurement theory applies to the measurement of canonical variables. Velocity, however, is a non canonical variable, since it is defined by Hamilton's equation rather than as an independent degree of freedom.
Thus, the measurement of velocity involves the Hamiltonian
\begin{equation}\label{eq:II:II:1}
 H'=\frac{p_{y}^{2}}{2m}-G(t)\dot{y}q=\frac{p_{y}^{2}}{2m}-G(t)\frac{\partial H'}{\partial p_{y}}q\,,
\end{equation}
where the gate function is defined by $G(t)=g_{0}t$ for $0\leq t\leq T_{0}$, $G(t)=G$  for $T_{0}\leq t\leq T $, $G(t)=G-g_{0}(t-T)$ for $T\leq t\leq T+T_{0} $ and $G(t)=0$ otherwise. $T\gg T_{0}$. This definition of the Hamiltonian seems to be circular. However, it is not. We have for the velocity
\begin{equation}\label{eq:II:II:2}
\dot{y}=\frac{p_{y}}{m}-G(t)\frac{\partial^{2} H'}{\partial p_{y}^{2}}q\,.
\end{equation}
Since all non relativistic Hamiltonians, including the unknown $H'$, are of second order in $p_{y}$ at most, the term involving $\frac{\partial H'}{\partial p_{y}}$ in
$(\ref{eq:II:II:1})$ is of first order in $p_{y}$ at most. Thus $\frac{\partial^{2} H'}{\partial p_{y}^{2}}=\frac{1}{m}.$ Note that this result is independent of the (possible) presence of an external vector potential, which has therefore been omitted. We obtain $\dot{y}=\frac{p_{y}}{m}-\frac{G(t)}{m}q$ and $H'=\frac{p_{y}^{2}}{2m}-G(t)\frac{p_{y}}{m}q+\frac{G^{2}(t)q^{2}}{m}$.
However, with this Hamiltonian we have for the pointer $ \dot{\pi}=-\frac{\partial H'}{\partial q}=G(t)\frac{p_{y}}{m}-\frac{2G^{2}(t)q}{m}$.
Since we want to measure the observed particle's original, unperturbed velocity, the $q$ dependence is undesirable. It may be eliminated by adding a compensatory interaction. We finally obtain,
\begin{equation}\label{eq:II:II:3}
   H = \frac{ \left[p_{y}-G(t)q \right]^{2} }{2m}-\frac{G^{2}(t)q^{2}}{2m}\,,
\end{equation}
where now $\delta\pi(t\geq T)=\int_{0}^{T}G(t)\frac{p_{y}}{m}dt=GT\frac{p_{y}(0)}{m}$, as desired.
\subsection{Using two position measurements}
\label{sec:2.3}
Below, it is shown that it is possible to measure velocity by means of two position measurements.
The Hamiltonian is given by
\begin{equation}
H=\frac{p_{y}^{2}}{2m}+g(t)qy-g(t-T)qy-\delta(t-T)\frac{a q^{2}}{2}\,.
\label{eq:II:III:1}
\end{equation}
The two position measurements are separated by an interval $T\gg T_{0}$. $\delta(t-T)$ is a Dirac $\delta$ function. The last term is the compensation term.
The kinetic energy of the measuring device has been ignored also in between the two position measurements. This entails that
$\frac{G^{2}T}{M}\rightarrow 0$.
Following the discussion in section $(\ref{sec:2.1})$, the particle's kinetic energy may be ignored during the interaction times $[0,T_{0}]$, $[T,T+T_{0}]$.
Thus, after the measurement, the pointer's displacement is given by
\begin{eqnarray}
  \delta\pi(t\geq T+T_{0}) &=& G[y(T)-y(0)]+aq(0) \nonumber \\
   &=& G\frac{p_{y}(0)-G q(0)}{m}T+aq(0) \nonumber \\
   &=& GTv_{y}(0)\; \Leftrightarrow\; a=\frac{G^{2}T}{m}\,.\label{eq:II:III:2}
\end{eqnarray}
With the compensation term properly tuned, the measuring device records \emph{the distance the particle would have traveled if it were not perturbed by the measurement}. In other words, it measures the particle's original velocity. Following $(\ref{eq:II:III:2})$ we assume $GT\sim 1$.
\paragraph{ }
After the second "hit", with a reversed force, the particle's velocity
returns to its initial value,
but with a greatly reduced uncertainty
$\Delta v_{y}(t\geq T+T_{0})=\frac{\Delta\pi(t\geq T+T_{0})}{GT}=\frac{\Delta\pi(0)}{GT}$, yet the product of the uncertainties remains
\begin{eqnarray}
 & &\Delta p_{y}(t\geq T+T_{0})\Delta y(t\geq T+T_{0}) \approx\nonumber\\
  & & m\frac{\Delta\pi(0)}{GT}\frac{G\Delta q(0)}{m}T
                                                          =\Delta\pi(0)\Delta q(0)\geq\hbar\,. \label{eq:II:III:3}
\end{eqnarray}
After the measurement, the uncertainty of the particle's position is greatly increased. This is since the particle moves at an uncertain velocity during the measurement. Thus, it is the uncertain vector potential appearing during the measurement, that "saves" the uncertainty principle from being violated.
\paragraph{ }
Finally, note that the general measurement of velocity involving the time dependent vector potential, arrived at in section $(\ref{sec:2.2})$, is equivalent, up to a gauge transformation, to that involving two position measurements described here. The pertinent gauge transformation is $U=e^{-iG(t)yq}$.
Thus, all one dimensional von Neumann velocity measurements are equivalent to a two-positions-measurement, and share the same characteristic features, namely the reversed forces effective at the beginning and the end and an uncertain velocity in between.
\section{The case of the AB effect}
\label{sec:3}
 I now relate specifically to the measurement of velocity in the context of the AB set-up. Note that, in order to obtain the velocity distribution, all one needs to do is to repeat
the measurement of the velocity for a large number of times $N$, and count the number $N_{s}$ of times the result $v_{s}$ is obtained, i.e.,
$P(v_{s})=\lim_{N\rightarrow\infty}\frac{N_{s}}{N}$.
\paragraph{ }
Initially, the charged particle is in a superposition of two wave packets
$\Psi(x,y)=\Psi_{1}+\Psi_{2}$
separated by a distance
$\vec{L}=L\vec{j}$, $L\gg \Delta y$, the width of the packets, and moving with velocity $\vec{v}_{o}=v_{o}\vec{i}$.
The packets' respective position distributions are equal, except for a displacement by $L$, i.e., $P_{2}(x,y)=P_{1}(x,y-L)$, where $P_{i}(x,y)=\Psi_{i}^{*}\Psi_{i}(x,y), i=1,2$.
The Fourier transform of the distribution of the velocity in the y direction, $P(v_{y})$, can be written as $f_{l}=\int
e^{imv_{y}l}P(v_{y})dmv_{y}=\langle e^{imv_{y}l}\rangle$, where $mv_{y}=p_{y}-\frac{e}{c}A_{y}(x,y)$, $p_{y}$ is the particle's momentum and $A_{y}$ is the vector potential due to the flux line. In reference \cite{Aharonov:Kaufherr} it has been shown that when the line connecting the wave packets crosses the flux line, the Fourier transforms for $l=\pm L$ change by
\begin{equation}\label{eq:III:1}
\delta f_{\pm L}=\delta \langle e^{\pm imv_{y}L}\rangle=\frac{1}{2}(e^{\pm i\alpha}-1),
\end{equation}
where $\alpha=-\frac{e\Phi}{\hbar c}$, $\Phi$ is the magnetic flux in the $+z$ direction and $e$ is the charge.
The velocity distribution is given by $P(v_{y})=\int dl e^{-imv_{y}l } f_{l}=\int dl e^{-imv_{y}l }\langle e^{imv_{y}l}\rangle$.
Using (\ref{eq:III:1}), one obtains that on passing by the flux line, it changes. Thus,
\begin{eqnarray}
  P_{before}(v_{y}) &=& 2P_{o}(v_{y})\cos^{2}\frac{mv_{y}L}{2\hbar} \label{eq:III:2}\,,\\
  P_{after}(v_{y}) &=& 2P_{o}(v_{y})\cos^{2}\frac{(mv_{y}-\alpha\frac{h}{L}  )L}{2\hbar}\,, \label{eq:III:3}
\end{eqnarray}
where $P_{o}(v_{y})$ is the velocity distribution of a single wave packet.
Equation (\ref{eq:III:1}) implies that the evolution of the charged particle in the vicinity of the flux line is essentially a one dimensional problem. In particular,  the velocity distributions  $(\ref{eq:III:2})$, $(\ref{eq:III:3})$, together with position distribution
$P'(y)=\int dx[P_{1}(x,y)+P_{2}(x,y)]$
 determine the states up to a choice of gauge. We choose:
\begin{eqnarray}
  \Psi_{before}(y,t)&\equiv &\Psi(y,t)\nonumber\\
  &=&\frac{1}{\sqrt{2}}[\psi(y+\frac{L}{2},t)+\psi(y-\frac{L}{2},t)]  \label{eq:III:4},\\
  \Psi_{after}(y,t)&\equiv&\Psi_{\alpha}(y,t)\nonumber\\
  &=&\frac{1}{\sqrt{2}}[\psi(y+\frac{L}{2},t)+e^{i\alpha}\psi(y-\frac{L}{2},t)] \label{eq:III:5},
\end{eqnarray}
where $\psi(y\pm\frac{L}{2})=\sqrt{P'}(y\pm\frac{L}{2})$ for $y\stackrel{\textstyle{<}}{>}0$.
Henceforth, the evolution of the charged particle is given in terms of these two one-dimensional states.
\paragraph{}We now turn to the measurement of velocity.
In the present context we shall use the Schr$\ddot{o}$dinger picture, where the resolution of the velocity paradox mentioned in the introduction becomes intuitively clear. Let us consider the particle's evolution. We have,
\begin{equation}\label{eq:III:6}
\left|\Psi_{s}(y,\pi,T)\right\rangle =\mathcal{T}e^{-i\int_{0}^{T}H(t)dt}\left|\Psi_{s}(y,\pi,0)\right\rangle,
\end{equation}
where $\left|\Psi_{s}(y,\pi,t)\right\rangle$ is the state of the entire system, i.e., measuring device and measured particle. $\mathcal{T}$ stands for time ordering. The Hamiltonian is given by $(\ref{eq:II:III:1})$.
We obtain\cite{appendix III:1}
\begin{eqnarray}
 \left|\Psi_{s}(y,\pi,T)\right\rangle&=&e^{i\frac{a q^{2}}{2}}e^{iGqy}e^{-i\frac{p_{y}^{2}}{2m}T}e^{-iGqy} \left|\Psi_{s}(y,\pi,0)\right\rangle \label{eq:III:7}\\
  &=&e^{-i\frac{p_{y}^{2}}{2m}T}e^{iG(\frac{p_{y}}{m}T)q} \left|\phi(\pi)\psi^{'}(y,0)\right\rangle \label{eq:III:8}\\
  &=&e^{iG(\frac{p_{y}}{m}T)q} \left|\phi(\pi)\psi^{'}(y,T)\right\rangle\label{eq:III:9}\\
  &=&\left|\phi(\pi-GTv_{y})\psi^{'}(y,T)\right\rangle \,, \label{eq:III:10}
\end{eqnarray}
provided $a=\frac{G^{2}T}{m}$. $\phi(\pi)$ and $\psi^{'}(y,t)$ are the wave functions of the measuring device and the measured particle, respectively. Since we are interested in the state of the particle after it has passed by the flux line, we substitute $\psi^{'}(y,t)=\Psi_{\alpha}(y,t)$. Equation $(\ref{eq:III:10})$ describes the state of the system after the measurement. The pointer is displaced by an amount that is proportional to the particle's velocity.
However, according to the physical interpretation of quantum theory, with the observed system in a state $\left|\psi^{'}\right\rangle$, a measurement of an observable $\hat{O}$ results, with probability
$|\langle o\left|\psi^{'}\right\rangle|^{2}$, in one of its eigenvalues $o$, with the observed system jumping into the corresponding eigenstate  $\left|o\right\rangle$. Thus, with a measurement of the particle's velocity, we have to expand its state
in eigenstates of velocity,
\begin{eqnarray}
 \left|\psi^{'}(y,T)\right\rangle&=&\left|\Psi_{\alpha}(y,T)\right\rangle\nonumber\\
 &=&\int f_{\Psi_{\alpha}}(mv_{y})e^{imv_{y}y}|mv_{y}>dmv_{y} \label{eq:III:11}\\
 &=&\sqrt{2}\int f_{\psi}(mv_{y})\cos\frac{(mv_{y}-\frac{\alpha}{L})L}{2} e^{imv_{y}y}|mv_{y}>dmv_{y}  \label{eq:III:12}
\end{eqnarray}
where $f_{\Psi_{\alpha}}$, $f_{\psi}$, are the Fourier transforms of the wave functions $\Psi_{\alpha}$ and $\psi$, respectively\cite{appendix III:2}. Substituting eqn. $(\ref{eq:III:12})$ into eqn. $(\ref{eq:III:10})$ we finally obtain
\begin{eqnarray}
\left|\Psi_{s}(y,\pi,T)\right\rangle&=&\sqrt{2}\int f_{\psi}(mv_{y})\cos\frac{(mv_{y}-\frac{\alpha}{L})L}{2} e^{imv_{y}y}\nonumber\\
& &\left|\phi(\pi-GTv_{y})\right\rangle|mv_{y}>dmv_{y}\,.\label{eq:III:13}
\end{eqnarray}
With the pointer variable $\pi$ sufficiently sharp, its uncertainty
\begin{equation}\label{eq:III:14}
\Delta\pi\ll \frac{GT}{m} \alpha\frac{h}{L}\leq \frac{GTh}{mL}\,,
\end{equation}
a sufficiently accurate measurement of the particle's velocity is possible. For an electron the change of velocity associated with modulo momentum, or the relative phase, is $\delta v_{y}=\frac{h}{mL}\approx 10^{4}\frac{cm}{sec}$, where we have taken $L\sim 6\mu m$ as in the late Professor Tonomura's sample\cite{Tonomura}.
>From $(\ref{eq:III:13})$ it follows that the pointer is displaced by $GTv_{y}$ with probability $2\mid f_{\psi}(mv_{y})\mid ^{2}\cos^{2}\frac{(mv_{y}-\frac{\alpha}{L})L}{2}$ . This is exactly the shifted velocity distribution $(\ref{eq:III:3})$ which we set out to measure. Thus, $(\ref{eq:III:13})$ and $(\ref{eq:III:14})$
demonstrate that it is possible to observe the shift of the velocity distribution affected by the flux line.
 It is also clear from equation $(\!\!~\ref{eq:III:13})$ that the measurement did not change the velocity distribution. This illustrates, in the Schr$\ddot{o}$dinger picture, the statement that a measurement of an observable $\hat{O}$ should not change it, namely, that it should not change its probability distribution. Clearly, a  measurement of velocity could not change any function of velocity such as the projection operator to one of its eigenvalues, regardless of the method by which the velocity is measured.
\paragraph{}Note that in the case when there is a relative velocity between the wave packets, as with the superposition $\Psi_{v_{o},\alpha}(y,t)=\frac{1}{\sqrt{2}}[\psi(y+\frac{L}{2},t)e^{iv_{o}y}+e^{i\alpha}\psi(y-\frac{L}{2},t)e^{-iv_{o}y}]$, the  velocity distributions of the two wave packets do not overlap and no shift of the velocity distribution is observed. However, by applying a suitable impulse to the wave packets the relative velocity may be annulled. The ensuing velocity distribution will display the $\alpha$-dependent shift. Repeating the experiment \emph{without} a flux line will prove that the relative phase existed in the original state, before the forces were applied.
\paragraph{}Note also that there is a relativistic limitation on the measurement of instantaneous velocity in the AB case.
Since the particle's uncertain intermediate velocity cannot exceed $c$, it follows from equation $(\!\!~\ref{eq:III:14})$ that
\begin{equation}\label{eq:III:II:15}
 c \geq G\frac{\Delta q}{m}\gg \frac{L}{T}\,,
 \end{equation}
 This is a limitation on measurements in the modulo region.
\paragraph{}To conclude: In the AB effect, the change of the velocity distribution of the charged particle occurs instantly. However, because of the separation $L$ between the wave packets, the duration of the \emph{measurement} of the particle's velocity is bounded from below by $T\gg \frac{L}{c}\sim 10^{-14}$sec. This gap in time is responsible for a discrepancy which exists in relativistic quantum theory, between what the theory predicts and what is observable. However, for $T\ll T_{travel}$ ($T_{travel}$ is the travel-time of the particle throughout the AB experiment) and, in particular, by varying $T_{travel}$ while keeping $T$ constant, the abrupt change of the velocity distribution \emph{in the vicinity of the flux line} may be convincingly demonstrated.
\section{Two Gedanken experiments}
\label{sec:4}
Consider the AB set-up in the charged particle's rest frame, where the flux line moves with velocity -$v_{0}\vec{i}$. In this reference frame the particle experiences only an electric AB effect, with the electric flux being equal to the magnetic flux in the flux line's rest frame.  The particle's velocity is measured twice, before the flux line has crossed the line connecting the wave packets, and immediately afterwards. Two position measurements involving the von Neumann coupling $g(t)qy$ and a compensation are used to measure velocity. Thus the abrupt change of the velocity distribution in the AB effect may be experimentally tested.
\paragraph{}Alternatively, the gauge invariant non local exchange may be tested by the interference experiment described in figure 2. Here, again, a typical AB set-up, except for a certain modification: Initially, no flux is threading the area enclosed by the (future) trajectories of the two wave packets. Shortly before the wave packets are about to pass by, at a certain $x=x_{0}$, say, the flux line is inserted into that area. Immediately after the
 particle has  passed by the flux line, the latter is removed again.
Thus, when the wave packets are brought together to interfere, there is no flux within the closed loop outlined by the wave packets. It seems that $\oint \vec{A}\cdot\vec{dl}=0$, and one may be led to expect no shift of the interference pattern.
However, since the line connecting the wave packets did cross the flux line before it was removed, the exchange must have occurred. The removal of the flux line does not undo this. Thus, a shift of the interference pattern should occur.
This prediction, if experimentally confirmed, demonstrates that the AB effect is contingent on the gauge invariant non local exchange, and on it alone.
\begin{figure}
\includegraphics[width=14.0cm]{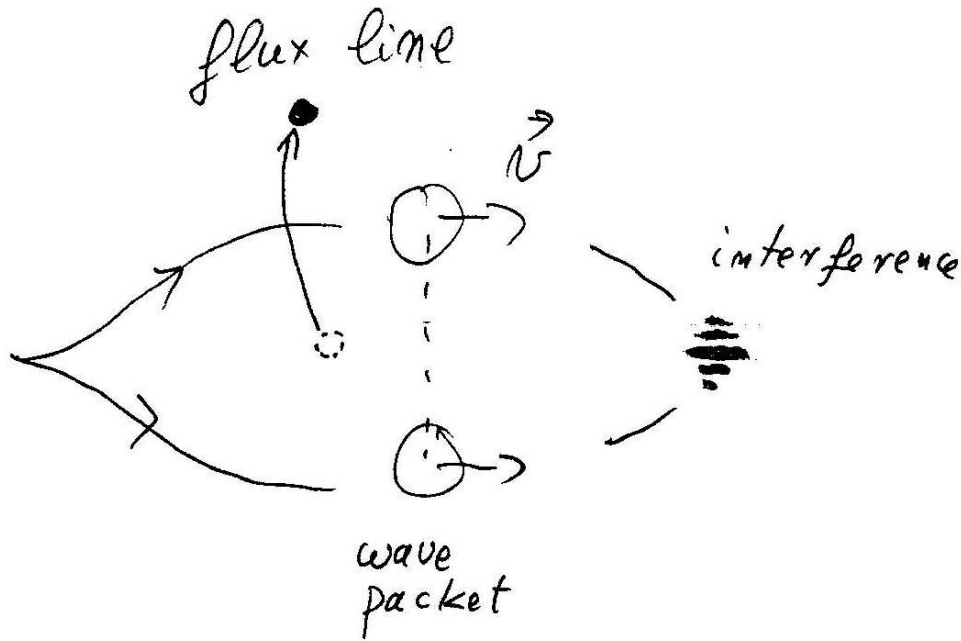}
\caption{As soon as the particle has passed by the flux line, the latter is removed.}
\end{figure}
\section{Conclusion}
\label{5}
It is shown in this paper that it is possible to verify experimentally the sudden shift of the velocity distribution occurring non locally in
 the AB effect. New light is shed on measurement of velocity in quantum theory.
\section{Acknowledgement}
Thanks to professor Yakir Aharonov for many useful discussions. Thanks also to professor Moti Segev for the
collaborative effort and support through an ERC grant. Support by ISF grant no. 1125/10 is acknowledged.


\begin{thebibliography}{999}
\bibitem{appendix III:1} 
In the Heisenberg representation we have
\begin{eqnarray}
  & &\mathcal{T}e^{-i\int_{0}^{T}H(t)dt}=\nonumber\\
  &=&e^{i\frac{a q^{2}}{2}}e^{iGqy}e^{-i\frac{p_{y}^{2}}{2m}T}e^{-iGqy}
  \label{eq:apendix III.1:1}\\
       &=&e^{i\frac{a q^{2}}{2}}e^{-i\frac{p_{y}^{2}}{2m}T}e^{i\frac{p_{y}^{2}}{2m}T}e^{iGqy}e^{-i\frac{p_{y}^{2}}{2m}T}e^{-iGqy}
       \label{eq:apendix III.1:2}\\
       &=&e^{-i\frac{p_{y}^{2}}{2m}T}e^{i\frac{a q^{2}}{2}}e^{i\frac{p_{y}^{2}}{2m}T}e^{iGqy}e^{-i\frac{p_{y}^{2}}{2m}T}e^{-iGqy}
       \label{eq:apendix III.1:3}\\
       &=&e^{-i\frac{p_{y}^{2}}{2m}T}e^{i\frac{a q^{2}}{2}}e^{iGq(y+\frac{p_{y}}{m}T)}e^{-iGqy}
       \label{eq:apendix III.1:4}\\
       &=&e^{-i\frac{p_{y}^{2}}{2m}T}e^{i\frac{a q^{2}}{2}}
       e^{-i\frac{(Gq)^{2}T}{2m}}e^{i\frac{GqT}{m}p_{y}}e^{iGqy}e^{-iGqy}
       \label{eq:apendix III.1:5}\\
       &=&e^{-i\frac{p_{y}^{2}}{2m}T}e^{i\frac{a q^{2}}{2}}e^{-i\frac{(Gq)^{2}T}{2m}}e^{i\frac{GqT}{m}p_{y}}
       \label{eq:apendix III.1:6}\\
       &=&e^{iG(\frac{p_{y}}{m}T)q}e^{-i\frac{p_{y}^{2}}{2m}T}
       \label{eq:apendix III.1:7}
\end{eqnarray}
For proof of equation $(\ref{eq:apendix III.1:4})$, note that in the Heisenberg representation we have $e^{i\frac{p_{y}^{2}}{2m}T}e^{iGqy}e^{-i\frac{p_{y}^{2}}{2m}T}=e^{iGqy(T)}$,
where the Heisenberg operator $y(T)$ is given by
$y(T)=e^{i\frac{p_{y}^{2}}{2m}T} y e^{-i\frac{p_{y}^{2}}{2m}T}=y+\frac{p_{y}}{m}T$.
Equation $(\ref{eq:apendix III.1:5})$ follows by using the
Baker Hausdorf formula $e^{i(\alpha p_{y}+\beta y)}=e^{-i\frac{\alpha\beta}{2}}e^{i\alpha p_{y}}e^{i\beta y}$ in $(\ref{eq:apendix III.1:4})$. Finally, by choosing $a=\frac{G^{2}T}{m}$, equation $(\ref{eq:apendix III.1:7})$, or $(\ref{eq:III:10})$, follows.
\bibitem{appendix III:2} 
The Fourier transform of a single wavepacket is
\begin{equation}\label{eq:III:apendix III.2:1}
  f_{\psi}(mv_{y})=\int dy e^{-imv_{y}y}\psi(y).
\end{equation}
The Fourier transform of the superposition $\Psi_{\alpha}$ is
\begin{equation}\label{eq:III:apendix III.2:2}
f_{\Psi_{\alpha}}(mv_{y})=\int dy e^{-imv_{y}y}\Psi_{\alpha}(y)=\sqrt{2}f_{\psi}(mv_{y})\cos\frac{(mv_{y}-p_{\alpha})L}{2},
\end{equation}
where $p_{\alpha}=\alpha\frac{h}{L}\equiv\frac{\alpha}{L}$. Note that the  Fourier transforms of the two orthogonal wavepackets are, of course, not orthogonal. However, they differ by the relative phase factor $e^{imv_{y}L}$. This is the origin of the cosine factor, or the shift of the velocity distribution. Finally, $P_{after}(mv_{y})=f_{\Psi_{\alpha}}^{*}f_{\Psi_{\alpha}}(mv_{y})$.
\end{thebibliography}
\end{document}